# BFRnet: A deep learning-based MR background field removal method for QSM of the brain containing significant pathological susceptibility sources

Xuanyu Zhu, Yang Gao, Feng Liu, Stuart Crozier, *Hongfu Sun

School of Information Technology and Electrical Engineering,

University of Queensland, Brisbane, Australia

*Correspondence: Hongfu Sun

Address: Room 540, General Purpose South (Building 78),

University of Queensland, St Lucia QLD 4072, Australia

Email: hongfu.sun@uq.edu.au

**Abstract**

**Introduction:** Background field removal (BFR) is a critical step required for successful quantitative susceptibility mapping (QSM). However, eliminating the background field in brains containing significant susceptibility sources, such as intracranial hemorrhages, is challenging due to the relatively large scale of the field induced by these pathological susceptibility sources.

**Method:** This study proposes a new deep learning-based method, BFRnet, to remove the background field in healthy and hemorrhagic subjects. The network is built with the dual-frequency octave convolutions on the U-net architecture, trained with synthetic field maps containing significant susceptibility sources. The BFRnet method is compared with three conventional BFR methods and one previous deep learning method using simulated and *in vivo* brains from 4 healthy and 2 hemorrhagic subjects. Robustness against acquisition field-of-view (FOV) orientation and brain masking are also investigated.

**Results:** For both simulation and *in vivo* experiments, BFRnet led to the best visually appealing results in the local field and QSM results with the minimum contrast loss and the most accurate hemorrhage susceptibility measurements among all five methods. In addition, BFRnet produced the most consistent local field and susceptibility maps between different sizes of brain masks, while conventional methods depend drastically on precise brain extraction and further brain edge erosions. It is also observed that BFRnet performed the best among all BFR methods for acquisition FOVs oblique to the main magnetic field.

**Conclusion**: The proposed BFRnet improved the accuracy of local field reconstruction in the hemorrhagic subjects compared with conventional BFR algorithms. The BFRnet method was effective for acquisitions of titled orientations and retained whole brains without edge erosion as often required by traditional BFR methods.

# 1. Introduction

In recent years, an MRI post-processing technique, namely quantitative susceptibility mapping (QSM) (1), has been developed for measuring magnetic susceptibility distribution *in vivo* (2). QSM application in the brain (3) has generated significant interest and has been investigated in multiple clinical studies, such as Parkinson's disease (4-7), Huntington's disease (7), Alzheimer's disease (8, 9), and healthy aging (10).

The measured magnetic field perturbation induced by a susceptibility distribution equals the convolution of this susceptibility with the unit dipole kernel (11). An ill-posed dipole deconvolution needs to be performed to recover the susceptibility source. However, the magnetic field outside the brain cannot be measured due to the lack of MR signals. Therefore, magnetic fields originating from susceptibility sources outside the brain (i.e. background field) generally need to be removed (12) before local field dipole inversion. Background field removal is the critical pre-processing step in the conventional QSM reconstruction pipeline to generate local field maps without the superposition of the substantially larger background field (13, 14). It removes field contribution from sources apart from local tissues, including air-tissue interfaces (e.g., at the skull and sinuses) (15), chemical shifts (16), and radiofrequency receiver offsets (17). It is essential for the successful implementation of QSM (18-22).

Several solutions have been proposed to eliminate the background magnetic field. For instance, the Projection onto Dipole Field (PDF) (21) method projects the field inside the brain onto the exterior-cerebral space and fits a surrounding susceptibility sources distribution. Sophisticated Harmonic Artifact Reduction for Phase data (SHARP) (23) separates the background field by the mean value property of the harmonic background field, which is further improved by RESHARP (18) and VSHARP (24) methods with Tikhonov regularization and variable spherical kernel size. Another method removes the background field by solving Laplacian Boundary Value (LBV) (20), assuming simple boundary conditions. However, these conventional methods either require brain edge erosion or are inaccurate at the brain boundary. Moreover, the performance of these methods depends substantially on the degree of edge erosions, which requires careful parameter tuning for brain masking. Thus, conventional background field removal significantly hinders the application of cortical QSM. Finally, total field dipole inversion methods (25, 26) have been proposed to eliminate the background field removal step and recover local susceptibility sources directly from the total field maps. However, these single-step methods do not produce the local field maps required by other methods such as COSMOS (27).

Recently, a deep learning-based method, namely SHARQnet (28), was proposed as an alternative solution to conventional background field removal methods. SHARQnet trained a U-net (29) with a simple background susceptibility source outside a digital phantom brain. However, SHARQnet has only been trained and tested on healthy brains; hence, the performance is unknown for diseased brains. A recent improvement on the U-net has been proposed by replacing the conventional convolution with an Octave operation (30). Previous studies have investigated the octave convolution to solve the QSM dipole inversion (31, 32). This study proposes a Background Field Removal (BFR) neural network, namely BFRnet, based on the octave convolutional U-net. The BFRnet is designed to remove the background field from outside of the brain without suppressing the local field from large local susceptibility sources, such as intracranial hemorrhage. Furthermore, BFRnet is investigated for removing the background field in arbitrary acquisition orientations relative to the main field.

## 2. Methods:

### 2.1 Theory

The magnetic field induced by a susceptibility distribution is equal to the convolution of the susceptibility with a unit dipole field kernel.

$$\delta B = F^{-1}DF\chi, \tag{1}$$

$\delta B$ and $\chi$ are the induced field map and the corresponding susceptibility source. $F$ and $F^{-1}$ are the forward and inverse Fourier Transforms. $D$ is the dipole kernel expressed in k-space:

$$D = \frac{1}{3} - \frac{k_z^2}{k_x^2 + k_y^2 + k_z^2}, \tag{2}$$

where $k_x$, $k_y$ represent the x and y coordination in-plane and $k_z$ indicates the direction paralleled to the main field B$_0$.

In this study, the original dipole forward calculation (Eq. (1)) was modified to avoid the aliasing effect due to circular convolution, written as:

$$\delta B = P^{-1}F^{-1}DFP\chi, \tag{3}$$

where $P$ refers to a zero-padding operation and $P^{-1}$ indicates the centre cropping operation.

The total susceptibility distribution $\chi$ is composed of the local brain tissue susceptibility $\chi_L$ and the background susceptibility $\chi_B$:

$$\chi = \chi_L + \chi_B \tag{4}$$

Combining Eqs. (1) and (4), the total field as the superposition of the local field ($\delta B_L$) and background field ($\delta B_B$), can be expressed as:

$$\delta B = \delta B_L + \delta B_B = P^{-1}F^{-1}DFP(\chi_L + \chi_B) \tag{5}$$

## 2.2 Training Data

The full-size (matrix size 144×192×128) brain susceptibility maps were obtained from 96 *in vivo* subjects (1 mm isotropic at 3T) and reconstructed using a previously developed pipeline (26). The BFRnet was trained on 28,800 small patches (matrix size:

$64^3$) by cropping the full-size QSM images, with 300 patches cropped randomly from each full-brain QSM volume.

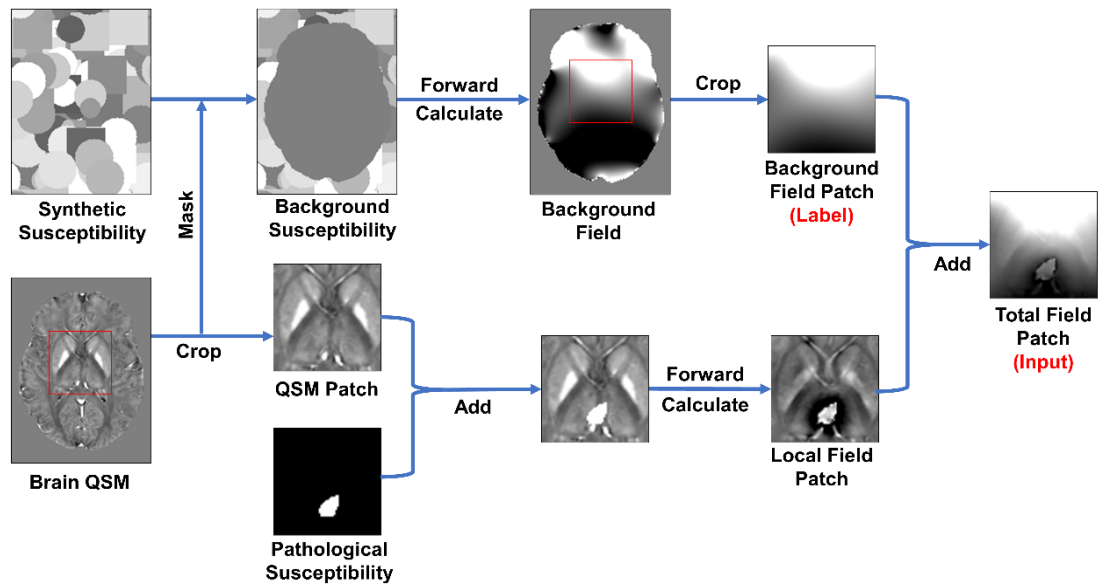

**Figure 1**. The generation pipeline of the BFRnet training dataset, including the background field and pathological susceptibility source simulation.

As illustrated in **Figure 1**, the background field was generated from simple synthetic susceptibility sources randomly distributed outside the region of interest (i.e., the full-brain image). A total of 600 susceptibility sources (i.e., cubes, spheres, and cylinders of various sizes) were randomly placed inside the entire image FOV. Background susceptibility source distribution was then simulated by masking the brain tissue region. Next, the background field was generated by the dipole field forward calculation, and patches of the size of $64^3$ were cropped from the full-size background field.

High susceptibility sources (e.g., hemorrhage and calcification) were simulated and included in the training set. Random shapes, including rectangles, cubes and spheres with 4 to 8 voxel diameters, were transformed by morphological operations (e.g., image dilation, erosion, opening and closing) to form irregular shapes and represent pathological lesions. The susceptibility values of these simulated pathological sources were set to normal distributions with a mean of -300 ppb for calcification and 800 ppb for hemorrhage. The standard deviation of the two normal distributions was set to 500 ppb. These high susceptibility sources were superimposed onto each original QSM patch at random locations inside the brain tissue. The forward model was used to calculate the local field patch, and the total field patches were constructed by adding the local field patches and the corresponding background field patches.

**2.3 Network Structure**

BFRnet comprises down-sampling and up-sampling paths, also called encoding and decoding. The BFRnet framework contains 10 Octave convolution layers (kernel size: 3×3×3), 2 max-pooling layers (kernel size: 2×2×2), 2 Octave transposed convolution layers (kernel size: 2×2×2), and 12 batch normalization layers. Rectified Linear Unit (ReLU) is adopted as the activation function of the network. The Octave convolution is explicitly designed to ensure that information in both resolutions (i.e., high and low) feature groups communicate efficiently (31). The architecture of BFRnet is illustrated in **Figure 2**. The mean squared error was adopted as the loss function during the training. The training rate was set to $10^{-3}$. Each BFRnet training took about 63 hours (45 epochs) using 2 Tesla V100 GPUs with a minibatch size of 24. The networks were implemented using MATLAB 2019a; the trained networks and the source codes are available at https://github.com/sunhongfu/deepMRI/tree/master/BFRnet.

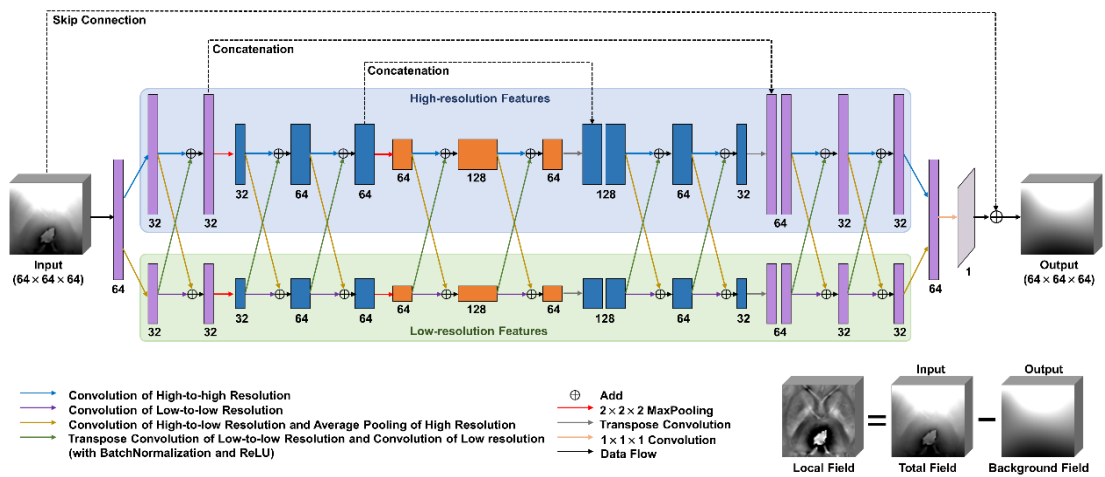

**Figure 2.** The BFRnet architecture, incorporating the octave convolution on the U-net architecture, with each selected feature divided into high- and low-resolution groups.

**2.4 Simulation and in vivo Experiments**

The COSMOS (27) acquisitions from one healthy human brain were used to generate three simulation datasets for testing. First, the PDF dipole fitting method (21) generated background susceptibility sources outside the brain. Then, the total susceptibility distribution was constructed by adding the COSMOS and background susceptibilities. The local and total field maps were simulated by the dipole forward calculations. For the second simulation test, we added a hemorrhage and calcification of random shapes with mean susceptibilities of 800 ppb and -400 ppb, respectively, onto the COSMOS maps to simulate a pathological dataset. We computed the local and total field maps for the third simulation test with an acquisition FOV 54-degree oblique from the main field $B_0$ on the healthy brain. The total field was fed as the network input for each testing dataset, and the simulated local field was regarded as the ground truth for evaluation.

For *in vivo* experiments, two total field maps (1 mm isotropic) were obtained by unwrapping the raw phase from 2 healthy subjects at 3 T. One of them was acquired with neutral orientation (pure axial), and the other has a 23-degree angle oblique from the main field direction ($B_0$). In addition, a hemorrhagic patient was scanned at 3 T with a pure axial acquisition, and the total field map was obtained after the Laplacian phase unwrapping (33).

The simulation experiments were performed on the full brains and *in vivo* tests were evaluated on both full-size and 3-voxel-edge-eroded brains. For comparison, three conventional BFR methods (e.g., PDF (21), LBV (20) and RESHARP (18)) and one deep learning-based method (SHARQnet (28)) were also performed.

The iLSQR (34) method was applied for dipole inversion to reconstruct QSM from all local field maps. For the three simulation tests, peak-signal-to-noise ratio (PNSR) and structural similarity (SSIM) relative to the ground truth were calculated for each method. For the two healthy *in vivo* experiments, susceptibility values of three deep grey matter (DGM) regions were measured, including globus pallidus (GP), putamen (PU), and caudate nucleus (CN). Moreover, the intracranial hemorrhage regions were segmented and measured, including simulation and the *in vivo* subjects. All ROIs were drawn on each subject's central three axial QSM slices, containing the largest areas of ROI regions. We measured these ROIs on results processed with full-brain and 3-voxel-erosion brain masks to investigate robustness against brain tissue masking.

## 3. Results:

### 3.1 Simulation Results

**Figure 3a** compares the local field results of different BFR methods on a simulated healthy subject, with the corresponding iLSQR-QSM shown in **Figure 3b**. BFRnet, PDF, and RESHARP tissue field results appear visually similar, while LBV appears dark and SHARQnet underestimates the tissue field contrast. After iLSQR dipole inversion, SHARQnet led to the most apparent errors, followed by RESHARP in the DGM regions. Quantitatively, BFRnet resulted in the second-best performance metrics with 48.19 dB of PSNR and 0.93 of SSIM, slightly lower than the PDF results (49.03 dB and 0.97). Similar trends were observed in susceptibility maps.

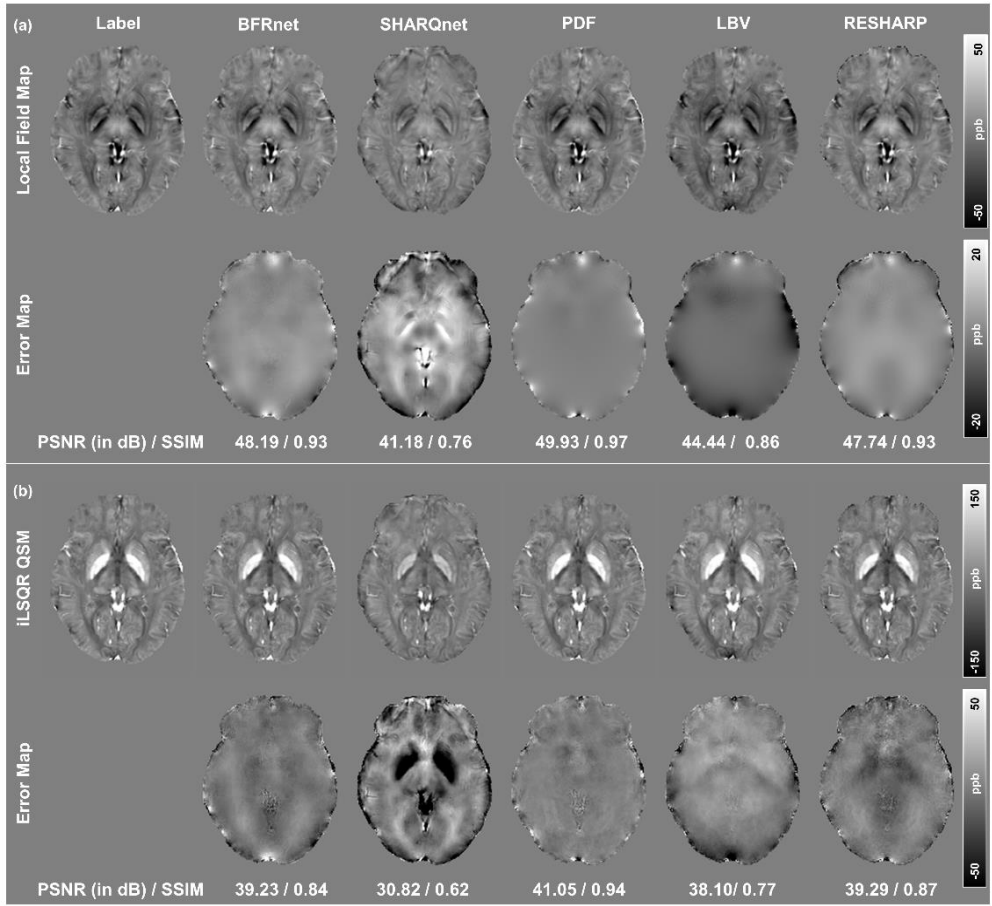

**Figure 3.** Comparison of BFRnet, SHARQnet, PDF, LBV and RESHARP on simulated brain data with the pure axial FOV orientation. The local field and difference maps to the ground truth are shown in **(a).** The corresponding QSM (by iLSQR) results are shown in **(b).**

**Figure 4** shows the BFR and QSM results of a simulation subject containing two high-susceptibility sources (i.e., hemorrhage and calcification). In local field results, the BFRnet showed the highest PSNR (48.30.40 dB) and SSIM (0.95), while SHARQnet showed the lowest PSNR (33.18 dB) and SSIM (0.67). The local field error maps confirmed that BFRnet led to the smallest error in high-susceptibility source regions, while all other methods, particularly SHARQnet, led to significant errors in the same regions. Consequently, these errors in local field maps resulted in substantial errors in the QSM results, as shown in **Figure 4b**. Measurements of the high-susceptibility pathological lesions are reported in **Table 1**. BFRnet achieved the most accurate susceptibility measurements followed by the LBV method, while SHARQnet led to substantial susceptibility underestimation in hemorrhage and calcification.

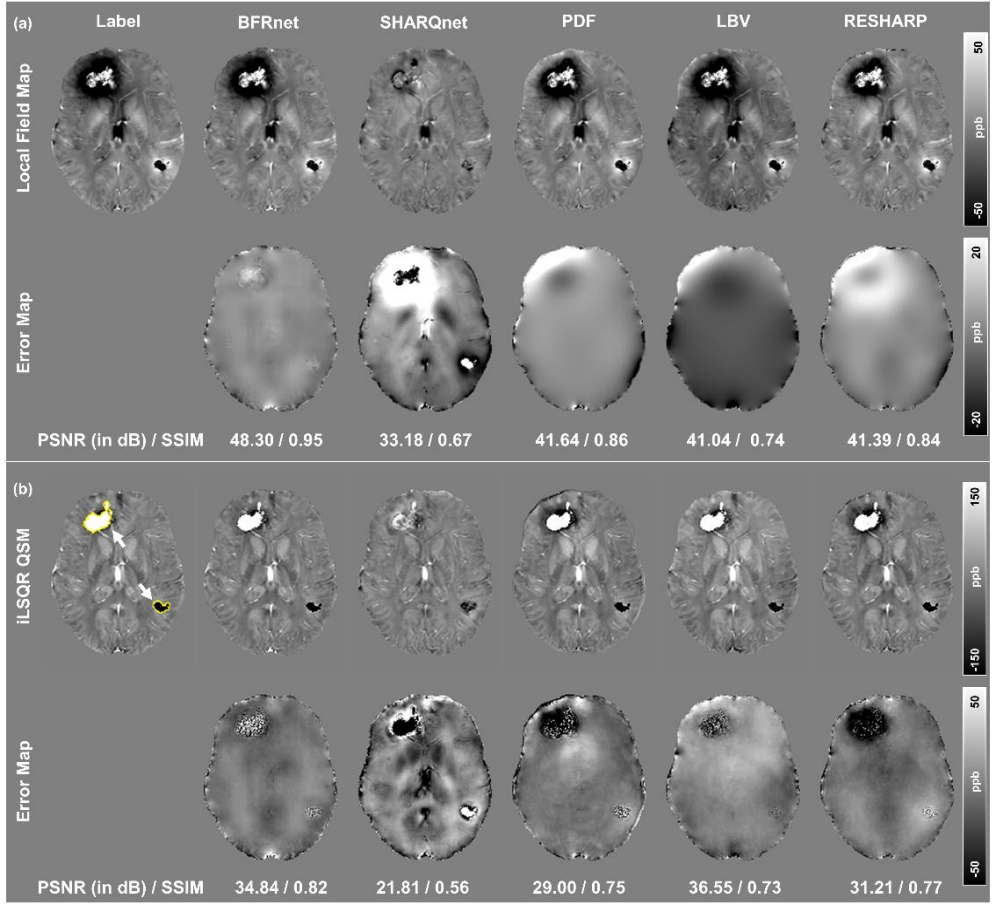

**Figure 4.** Local field **(a)** and iLSQR **(b)** results from the pathological simulation subject using five BFR methods. The measured pathological ROIs were outlined in yellow and pointed by the white arrows. Error maps relative to the ground truth are shown in the bottom rows of **(a)** and **(b)**.

**Table 1.** Susceptibility measurements (mean ± standard deviation) of simulated the simulated hemorrhage and calcification (unit: ppb)

|  | Label | BFRnet | SHARQnet | PDF | LBV | RESHARP |
|---|---|---|---|---|---|---|
| Hemorrhage | 632 ± 150 | **633 ± 150** | 103 ± 98 | 521 ± 161 | 611 ± 151 | 525 ± 127 |
| Calcification | -361 ± 99 | **-361 ± 101** | -122 ± 54 | -364 ± 105 | -371 ± 100 | -346 ± 101 |

The most accurate measurements from the five methods were highlighted in **bold**.

**Figure 5** demonstrates the performance of different BFR methods on the simulation subject with a tilted FOV of 54-degree to the main field. Similar to the pure axial simulation findings, the PDF method performed the best, followed by BFRnet. LBV showed apparent errors close to brain boundary, with the lowest PSNR of 37.59 dB. SHARQnet shows substantial contrast loss in the resulting local field map, with the lowest SSIM of 0.82. Reconstruction artifacts were also observed in SHARQnet, as pointed by the yellow arrow. Similar trends were observed in susceptibility maps using iLSQR for dipole inversion, as shown in **Figure 5b**.

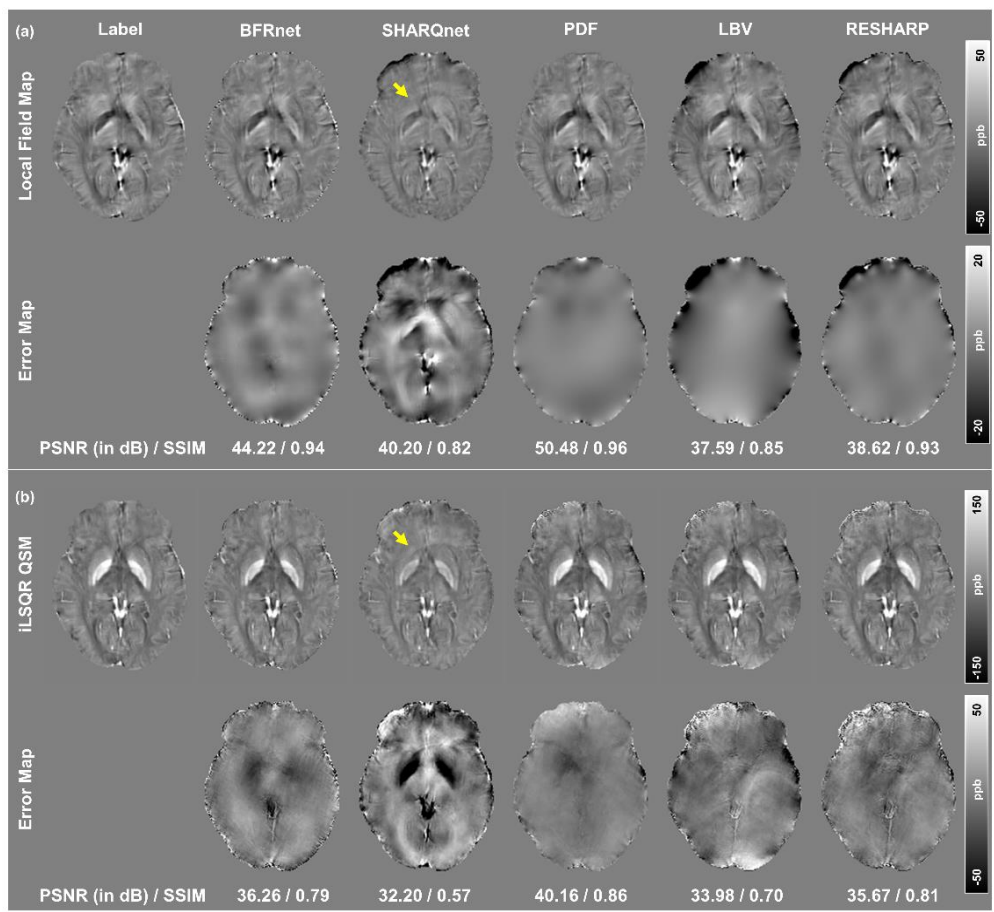

**Figure 5.** Comparison of five different BFR methods on a simulation subject with 54-degree oblique head orientation to the main field. The local field and QSM results are demonstrated in **(a)** and **(b)**. Error maps related to the labels are displayed at the bottom rows. The yellow arrows point to the local field and susceptibility artifacts in the SHARQnet results in **(b)**.

## 3.2 *In vivo* experiments

**Figure 6a** demonstrates the background field removal results from an *in vivo* acquisition with pure axial FOV, using full-brain and 3-voxel-erosion brain masks. The corresponding QSM results are shown in **Figure 6b**. The difference maps between full-brain and 3-voxel-erosion results are displayed in the third row. It can be seen that BFRnet and RESHARP resulted in the most visually appealing local field and susceptibility maps. The most substantial errors were observed in the LBV results using the full-brain mask. Deep learning-based methods (i.e., BFRnet and SHARQnet) gave the most consistent results between the two different brain masks due to the local convolution property (32). Susceptibilities of three DGM regions (i.e., GP, PU and CN) from different BFR methods were compared in **Figure 6c** from both full-brain and 3-voxel-eroded brain masks. RESHARP showed measurement variations of 30 ppb, 15 ppb and 16 ppb in GP, PU and CN; while LBV showed 26 ppb in GP, 23 ppb in PU and 13 ppb in CN, between the two brain masks. Meanwhile, measurement variations from different brain masking of the BFRnet method were much smaller, with 1 ppb in GP and 6 ppb in both PU and CN. Note that susceptibility measurements from SHARQnet are significantly underestimated compared to other methods.

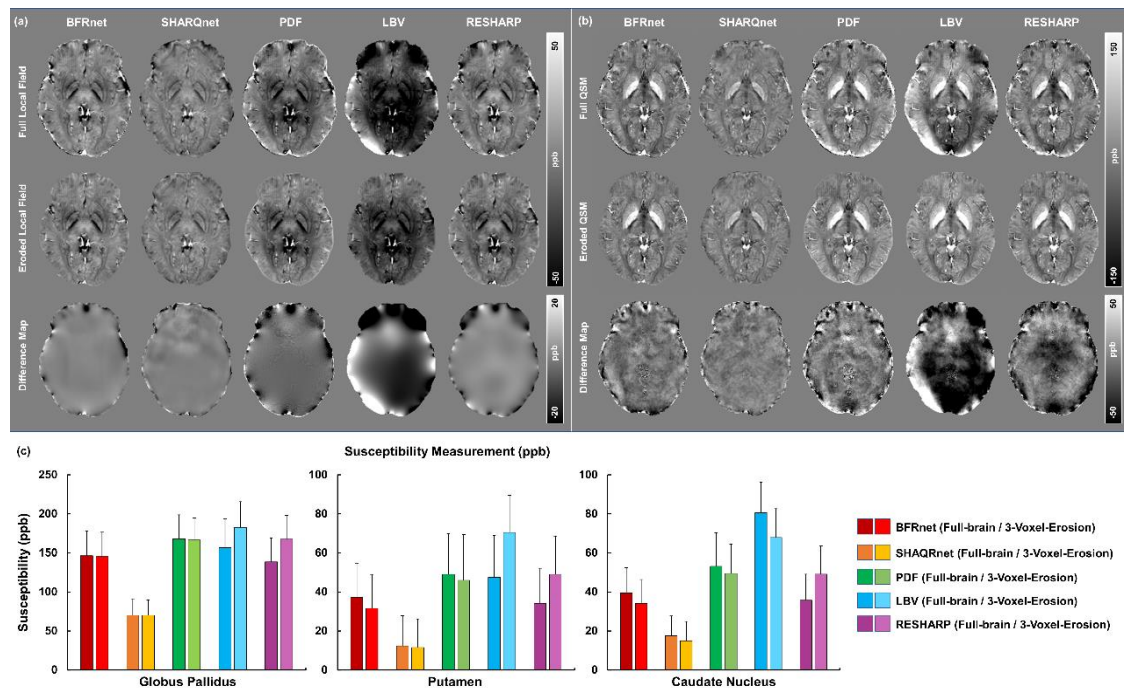

**Figure 6. (a)** Local field and **(b)** QSM results from an *in vivo* healthy subject in the pure axial orientation parallel B$_0$. The difference maps using full and 3-voxel-eroded brain masks are illustrated below. Measurements (mean and standard deviation) of three ROIs are reported in **(c)**, including globus pallidus, putamen, and caudate nucleus.

**Figure 7** demonstrates the background field removal and the corresponding QSM results of an *in vivo* acquisition from a hemorrhagic patient with the pure axial acquisition, using full-brain and 3-voxel-eroded brain masks. Similar to the findings from the healthy subject above, BFRnet and SHARQnet displayed fewer artifacts, while LBV, RESHARP and PDF results suffered from overwhelming artifacts. For quantitative analysis, **Table 2** reports the susceptibility measurements of the hemorrhage region, as outlined in yellow in **Figure 7b**. BFRnet exhibited the most consistent hemorrhage susceptibility measurements, with only a 4.6% difference between the full-brain and 3-voxel-eroded masks, compared to 27.2% for the LBV method. The hemorrhage measurements from BFRnet are consistent with the RESHARP results, while SHARQnet drastically underestimated the hemorrhage susceptibility more than 10 times compared to other methods.

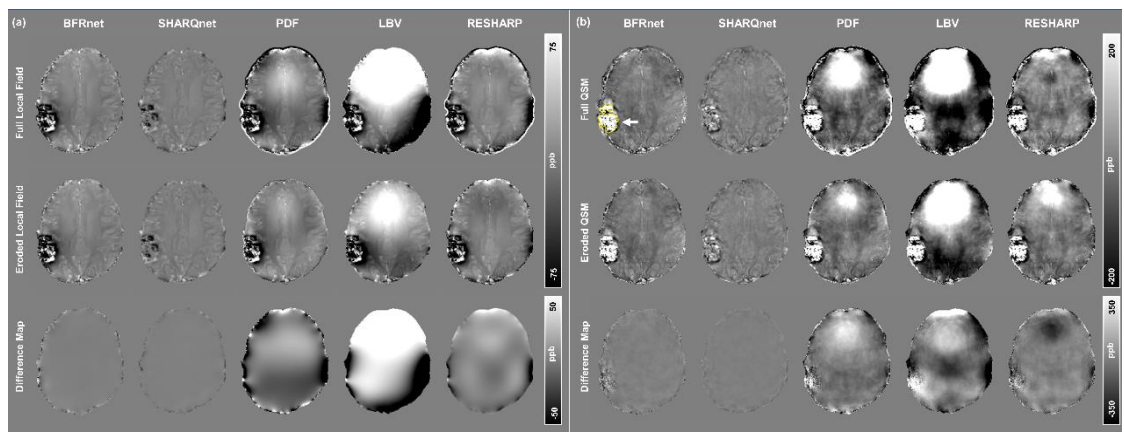

**Figure 7. (a)** Local field and **(b)** QSM results from an *in vivo* hemorrhagic patient. The measured hemorrhage ROI is outlined in yellow and pointed by the white arrow. The difference maps between full-brain and 3-voxel-eroded-brain results are shown in the last row.

**Table 2.** Susceptibility measurements (mean ± standard deviation) of the intracranial hemorrhage (shown in Fig. 7) from an *in vivo* patient experiment using two brain masks (unit: ppb)

|  | BFRnet | SHARQnet | PDF | LBV | RESHARP |
|---|---|---|---|---|---|
| Full-brain | 455 ± 331 | 43 ± 76 | 532 ± 350 | 639 ± 413 | 457 ± 353 |
| 3-voxel-erosion | 434 ± 325 | 41 ± 73 | 476 ± 349 | 465 ± 322 | 448 ± 340 |

**Figure 8** illustrates the *in vivo* experiment results of BFR methods on an oblique acquisition, with FOV 23-degree angled to the main field. The difference maps between full and eroded brain masks are illustrated in the bottom row. It is observed that the proposed BFRnet achieved the most visually appealing local field and susceptibility results, with no substantial artifacts in either local field or susceptibility results. It is also seen that deep learning-based methods (i.e., BFRnet and SHARQnet) produce the minimum variations between the full-brain and edge-eroded-brain reconstructions. On the contrary, PDF, LBV and RESHARP showed significant residual artifacts near the sinus region. Note that SHARQnet not only significantly suppressed local field and susceptibility contrasts but also introduced susceptibility artifacts near the internal capsule, as pointed by yellow arrows in **Figure 8b**. Susceptibility measurements of three DGM regions are plotted in **Figure 8c**. Similar to the pure axial acquisition experiments, BFRnet presented the most consistent DGM susceptibility measurements between two brain masks, while LBV displayed the highest measurement variations.

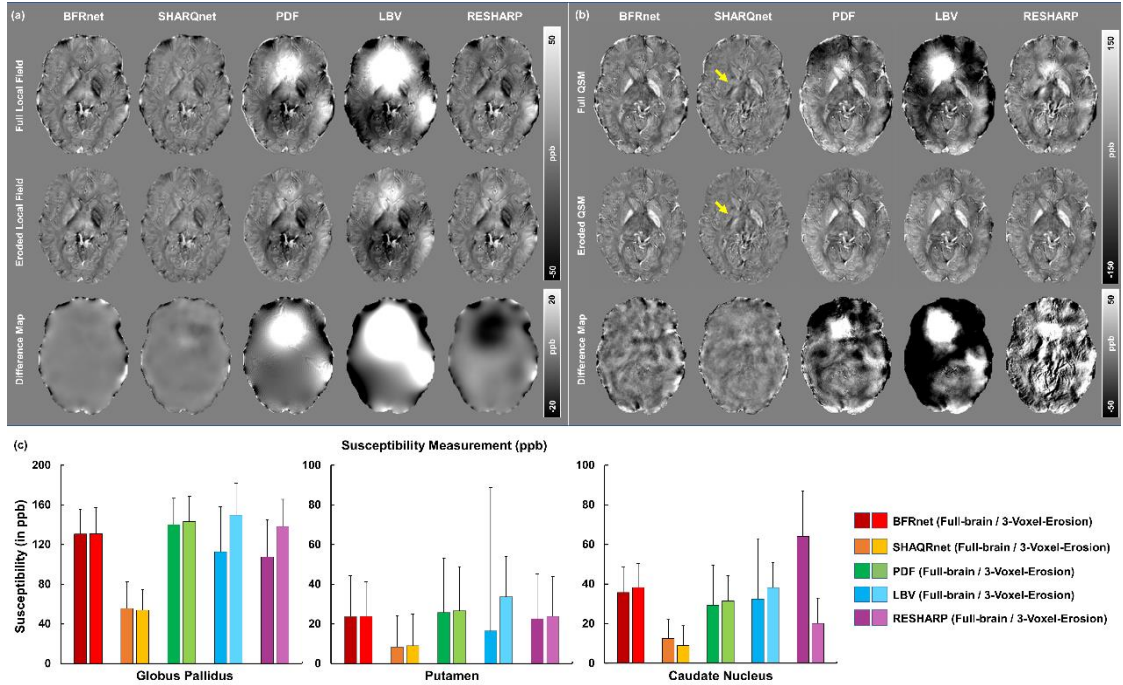

**Figure 8**. **(a)** Local field and **(b)** QSM results from an *in vivo* healthy subject with acquisition FOV 23-degree angled from the main field direction ($B_0$). The difference maps of the two brain masks are presented in the third row. Measurements of three DGM regions (globus pallidus, putamen, and caudate nucleus) are plotted in **(c)**. Yellow arrows point to susceptibility artifacts near the internal capsule in SHARQnet results.

## 4. Discussion

We proposed a learning-based method, BFRnet, to solve the problem of background field removal, particularly for cases of large susceptibility sources, such as hemorrhages, in the brain. Three total field maps were simulated through the forward model, and another three total field maps were obtained by post-processing (i.e., phase unwrapping) the raw phase maps from *in vivo* GRE acquisitions. Both simulation and *in vivo* experiments include one healthy subject, one hemorrhagic patient, and the same healthy subject but acquired in a tilted FOV. We evaluated the performance of BFRnet by standard error metrics (i.e., SSIM and PSNR) and mean susceptibilities of deep grey matter and hemorrhage regions. It is observed that the proposed BFRnet method achieved similar performances in healthy subjects and improved accuracy in pathological cases, compared with three conventional iterative algorithms (i.e., PDF, LBV, and RESHARP) and a previous deep learning method (i.e., SHARQnet), without the need for brain edge erosion. Our results also demonstrated that even though the BFRnet was trained on pure-axial acquisitions, it can be generalized to arbitrary FOV orientations relative to the main magnetic field direction.

Our proposed BFRnet shares some similarities with a previously published data-driven method SHARQnet. However, the more diverse training data and advanced network architecture significantly improved the BFRnet results. For example, each background field in SHARQnet was calculated by three background susceptibility sources outside of the brain. While in BFRnet, more than 600 random background sources were generated that covered the entire non-brain area inside FOV. Hence, the background field map in BFRnet delivered a more complex and realistic scenario than SHARQnet. In addition, BFRnet cropped patches from 96 different human brain volumes, whereas SHARQnet only utilized one simulated brain data. This enhancement further increases the complexity of the network training. These may explain why BFRnet is more robust than SHARQnet when FOV orientation is oblique to the main field, as shown in Figures 5 and 8. Moreover, simulated high-susceptibility sources were included in the local susceptibility maps to extend the local field value range during BFRnet training. This significantly boosted the performance of BFRnet on the diseased brain with hemorrhage inside, which introduces large-scale local field perturbation. On the other hand, these large-scale local field dipoles originated from pathologies are outside the range of SHARQnet since it has been trained on healthy brains only with substantially smaller local field variations from healthy brain tissue. Lastly, our BFRnet was built with the dual-frequency octave convolution, which benefits from low- and high-frequency spatial information interactions.

BFRnet also exhibited a high capability in pathological experiments with lower contrast loss and more precise susceptibility measurements than three conventional BFR methods. The well-designed training data with the simulated lesion regions boosted the network's adaptability in pathological cases and contributed to the successful reconstruction. Conventional methods failed pathological tests since part of the large-scale local field from pathological susceptibilities was also modelled and removed during the background field removal process. Therefore, these conventional background field removal methods underestimated the dipole from lesion regions. The error was further passed to the resulting QSM.

We also investigated the background field removal in oblique FOV acquisitions. It is found that the performance in oblique acquisitions was less robust than in pure axial acquisitions for all BFR methods. However, the proposed BFRnet performed significantly better than other methods, particularly for *in vivo* experiments, while PDF and LBV resulted in apparent residual background fields, and SHARQnet suppressed contrast and exhibited artifacts in QSM. Therefore, although pure axial acquisition data were used for training, BFRnet is generalized to arbitrary FOV orientations.

Moreover, it is observed that both learning-based methods (i.e., SHARQnet and BFRnet) are robust against brain masking. On the other hand, conventional methods are substantially affected by phase errors at the brain edge and require a precise extraction and further erosions of the brain. Significant artifacts were produced in conventional BFR methods when the full-size brains were kept, and these artifacts were substantially reduced after 3 voxels of edge erosion. In comparison, BFRnet did not produce noticeable artifacts in the full brain results and was consistent with brain edge eroded results. This enhanced performance originates from the architecture of the convolutional network. A small localized receptive field input was convoluted with the kernel for each convolutional layer. Thus, the errors at the edge of the brain only affect the local edge voxels but not the centre regions of the image (32). Therefore, BFRnet is helpful to solve the edge erosion problem in conventional BFR methods and can benefit applications of cortical QSM.

This study has a few limitations. First, we have only evaluated testing datasets with 1 mm$^3$ isotropic image resolution since this matches the training datasets. BFRnet's application on total field maps of other image resolutions may need retraining with datasets of matched image resolution. Second, only intracranial hemorrhages and calcifications were investigated as pathological brains for network evaluation. Diseased brains with other abnormalities have not been tested to evaluate the generalization of the BFRnet. Finally, BFRnet, as the second last step of the conventional QSM reconstruction pipeline, is affected by errors accumulated from previous processing steps, such as phase unwrapping (35, 36). Further study will investigate the feasibility of combining BFRnet with the previous processing steps to improve the accuracy of the entire QSM reconstruction pipeline.

## 5. Conclusion

A deep learning-based background field removal network, BFRnet, was proposed in this study, which improved the accuracy of local field reconstruction in the hemorrhagic subjects compared with conventional algorithms (i.e., PDF, LBV, and RESHARP) and a previous deep learning method (i.e., SHARQnet). The BFRnet also performs well in tilted acquisitions oblique to the main field. More importantly, BFRnet outputs local field maps of the full brain without edge erosion, which may enable the application of cortical QSM for clinical studies.

## 6. Acknowledgment

HS acknowledges support from the Australian Research Council (DE210101297).

## Reference

1. Haacke EM, Liu S, Buch S, Zheng W, Wu D, Ye Y. Quantitative susceptibility mapping: current status and future directions. Magnetic resonance imaging. 2015;33(1):1-25.

2. Sun H, Walsh AJ, Lebel RM, Blevins G, Catz I, Lu J-Q, et al. Validation of quantitative susceptibility mapping with Perls' iron staining for subcortical gray matter. Neuroimage. 2015;105:486-92.
3. Lim IAL, Faria AV, Li X, Hsu JT, Airan RD, Mori S, et al. Human brain atlas for automated region of interest selection in quantitative susceptibility mapping: application to determine iron content in deep gray matter structures. Neuroimage. 2013;82:449-69.
4. Langkammer C, Pirpamer L, Seiler S, Deistung A, Schweser F, Franthal S, et al. Quantitative susceptibility mapping in Parkinson's disease. PLoS One. 2016;11(9):e0162460.
5. Dexter D, Carayon A, Javoy-Agid F, Agid Y, Wells F, Daniel S, et al. Alterations in the levels of iron, ferritin and other trace metals in Parkinson's disease and other neurodegenerative diseases affecting the basal ganglia. Brain. 1991;114(4):1953-75.
6. Acosta-Cabronero J, Cardenas-Blanco A, Betts MJ, Butryn M, Valdes-Herrera JP, Galazky I, et al. The whole-brain pattern of magnetic susceptibility perturbations in Parkinson's disease. Brain. 2017;140(1):118-31.
7. Chen JC, Hardy P, Kucharczyk W, Clauberg M, Joshi J, Vourlas A, et al. MR of human postmortem brain tissue: correlative study between T2 and assays of iron and ferritin in Parkinson and Huntington disease. American journal of neuroradiology. 1993;14(2):275-81.
8. Bouras C, Giannakopoulos P, Good PF, Hsu A, Hof PR, Perl DP. A laser microprobe mass analysis of brain aluminum and iron in dementia pugilistica: comparison with Alzheimer's disease. European neurology. 1997;38(1):53-8.
9. LeVine SM. Iron deposits in multiple sclerosis and Alzheimer's disease brains. Brain research. 1997;760(1-2):298-303.
10. Betts MJ, Acosta-Cabronero J, Cardenas-Blanco A, Nestor PJ, Düzel E. High-resolution characterisation of the aging brain using simultaneous quantitative susceptibility mapping (QSM) and R2* measurements at 7 T. Neuroimage. 2016;138:43-63.
11. Polak D, Chatnuntawech I, Yoon J, Iyer SS, Milovic C, Lee J, et al. Nonlinear dipole inversion (NDI) enables robust quantitative susceptibility mapping (QSM). NMR in Biomedicine. 2020;33(12):e4271.
12. Schweser F, Deistung A, Reichenbach JR. Foundations of MRI phase imaging and processing for Quantitative Susceptibility Mapping (QSM). Zeitschrift für medizinische Physik. 2016;26(1):6-34.
13. Fortier V, Levesque IR. Phase processing for quantitative susceptibility mapping of regions with large susceptibility and lack of signal. Magnetic resonance in medicine. 2018;79(6):3103-13.
14. Özbay PS, Rossi C, Kocian R, Redle M, Boss A, Pruessmann KP, et al. Effect of respiratory hyperoxic challenge on magnetic susceptibility in human brain assessed by quantitative susceptibility mapping (QSM). NMR in Biomedicine. 2015;28(12):1688-96.

15. Buch S, Liu S, Ye Y, Cheng YCN, Neelavalli J, Haacke EM. Susceptibility mapping of air, bone, and calcium in the head. Magnetic resonance in medicine. 2015;73(6):2185-94.
16. Dimov AV, Liu T, Spincemaille P, Ecanow JS, Tan H, Edelman RR, et al. Joint estimation of chemical shift and quantitative susceptibility mapping (chemical QSM). Magnetic resonance in medicine. 2015;73(6):2100-10.
17. Levitt MH. The signs of frequencies and phases in NMR. Journal of Magnetic Resonance. 1997;126(2):164-82.
18. Sun H, Wilman AH. Background field removal using spherical mean value filtering and Tikhonov regularization. Magnetic resonance in medicine. 2014;71(3):1151-7.
19. Wang S, Liu T, Chen W, Spincemaille P, Wisnieff C, Tsiouris AJ, et al. Noise effects in various quantitative susceptibility mapping methods. IEEE Transactions on Biomedical Engineering. 2013;60(12):3441-8.
20. Zhou D, Liu T, Spincemaille P, Wang Y. Background field removal by solving the Laplacian boundary value problem. NMR in Biomedicine. 2014;27(3):312-9.
21. Liu T, Khalidov I, de Rochefort L, Spincemaille P, Liu J, Tsiouris AJ, et al. A novel background field removal method for MRI using projection onto dipole fields. NMR in Biomedicine. 2011;24(9):1129-36.
22. Groetsch C. The theory of tikhonov regularization for fredholm equations. 104p, Boston Pitman Publication. 1984.
23. Schweser F, Deistung A, Lehr BW, Reichenbach JR. Quantitative imaging of intrinsic magnetic tissue properties using MRI signal phase: an approach to in vivo brain iron metabolism? Neuroimage. 2011;54(4):2789-807.
24. Li W, Wu B, Liu C. Quantitative susceptibility mapping of human brain reflects spatial variation in tissue composition. Neuroimage. 2011;55(4):1645-56.
25. Liu Z, Kee Y, Zhou D, Wang Y, Spincemaille P. Preconditioned total field inversion (TFI) method for quantitative susceptibility mapping. Magnetic resonance in medicine. 2017;78(1):303-15.
26. Sun H, Ma Y, MacDonald ME, Pike GB. Whole head quantitative susceptibility mapping using a least-norm direct dipole inversion method. NeuroImage. 2018;179:166-75.
27. Liu T, Spincemaille P, De Rochefort L, Kressler B, Wang Y. Calculation of susceptibility through multiple orientation sampling (COSMOS): a method for conditioning the inverse problem from measured magnetic field map to susceptibility source image in MRI. Magnetic Resonance in Medicine: An Official Journal of the International Society for Magnetic Resonance in Medicine. 2009;61(1):196-204.
28. Bollmann S, Kristensen MH, Larsen MS, Olsen MV, Pedersen MJ, Østergaard LR, et al. SHARQnet–Sophisticated harmonic artifact reduction in quantitative susceptibility mapping using a deep convolutional neural network. Zeitschrift für Medizinische Physik. 2019;29(2):139-49.

29. Ronneberger O, Fischer P, Brox T, editors. U-net: Convolutional networks for biomedical image segmentation. International Conference on Medical image computing and computer-assisted intervention; 2015: Springer.
30. Chen Y, Fan H, Xu B, Yan Z, Kalantidis Y, Rohrbach M, et al., editors. Drop an octave: Reducing spatial redundancy in convolutional neural networks with octave convolution. Proceedings of the IEEE/CVF International Conference on Computer Vision; 2019.
31. Gao Y, Zhu X, Moffat BA, Glarin R, Wilman AH, Pike GB, et al. xQSM: quantitative susceptibility mapping with octave convolutional and noise-regularized neural networks. NMR in Biomedicine. 2021;34(3):e4461.
32. Zhu X, Gao Y, Liu F, Crozier S, Sun H. Deep grey matter quantitative susceptibility mapping from small spatial coverages using deep learning. Zeitschrift für Medizinische Physik. 2021.
33. Li W, Avram AV, Wu B, Xiao X, Liu C. Integrated Laplacian-based phase unwrapping and background phase removal for quantitative susceptibility mapping. NMR in Biomedicine. 2014;27(2):219-27.
34. Li W, Wang N, Yu F, Han H, Cao W, Romero R, et al. A method for estimating and removing streaking artifacts in quantitative susceptibility mapping. Neuroimage. 2015;108:111-22.
35. Abdul-Rahman HS, Gdeisat MA, Burton DR, Lalor MJ, Lilley F, Moore CJ. Fast and robust three-dimensional best path phase unwrapping algorithm. Applied optics. 2007;46(26):6623-35.
36. Robinson SD, Bredies K, Khabipova D, Dymerska B, Marques JP, Schweser F. An illustrated comparison of processing methods for MR phase imaging and QSM: combining array coil signals and phase unwrapping. NMR in Biomedicine. 2017;30(4):e3601.